\documentclass[10pt,twocolumn,amssymb,amsmath,aps,pra]{revtex4-1}
\usepackage{graphicx}

\newcommand*{\Scale}[2][4]{\scalebox{#1}{$#2$}}

\begin{document}

\title{Thermal balance and photon-number quantization in layered structures}
\date{\today}
\author{Mikko Partanen}
\author{Teppo H\"ayrynen}
\author{Jani Oksanen}
\author{Jukka Tulkki}
\affiliation{Department of Biomedical Engineering and Computational Science, Aalto University, P.O. Box 12200, 00076 Aalto, Finland}

\begin{abstract}
The quantization of the electromagnetic field in lossy and dispersive dielectric media has
been widely studied during the last few decades. However, several aspects of energy
transfer and its relation to consistently defining position-dependent ladder operators
for the electromagnetic field in nonequilibrium conditions have partly escaped
the attention. In this work we define the position-dependent ladder operators and an
effective local photon-number operator that are consistent
with the canonical commutation relations and use these concepts to describe the energy
transfer and thermal balance in layered geometries. This approach results in a position-dependent
photon-number concept that is simple and consistent with classical energy
conservation arguments. The operators are formed by first calculating the vector potential operator using 
Green's function formalism and Langevin noise source operators related to the medium and its 
temperature, and then defining the corresponding position-dependent annihilation operator
that is required to satisfy the canonical commutation relations in arbitrary geometry.
Our results suggest that the effective photon number associated with the electric field is generally
position dependent and enables a straightforward method to calculate the energy transfer rate
between the field and the local medium. In particular, our results predict that the
effective photon number in a vacuum cavity formed between two lossy material layers can
oscillate as a function of the position suggesting that also the local field temperature
oscillates. These oscillations are expected to be directly observable using relatively straightforward
experimental setups in which the field-matter interaction is dominated by the coupling to the electric field.
The approach also gives further insight on separating the photon ladder
operators into the conventional right and left propagating parts and on the anomalies
reported for the commutation relations of the corresponding operators within optical cavities.
\end{abstract}

\maketitle

\section{Introduction}
Recent research of nanoscale radiative energy transfer has enabled advances and in-depth insight in several
fields of optical technologies related, e.g., to nanoplasmonics \cite{Sorger2011,Oulton2009,Huang2013,Sadi2013},
near-field microscopy \cite{Taubner2006,Hillenbrand2002}, thin-film light-emitting diodes \cite{Nakamura2013,Heikkila2013},
photonic crystals \cite{Russell2003,Akahane2003}, and metamaterials \cite{Tanaka2010,Mattiucci2013}.
The optical phenomena taking place in nanoscale also naturally lead to questions related to the quantum
nature of light, proper ways to quantize the fields, and maybe even more importantly on how to
interpret the results. The research related to quantum optics of nanoscale systems is strongly
influenced by the availability of simple and transparent theoretical tools and
models that allow in-depth understanding of the pertinent phenomena
in sufficiently simple form. Such insight has been used, for instance, in the recent
demonstrations and studies of 
noiseless but nondeterministic optical amplifiers \cite{Zavatta2011,Ferreyrol2010,Partanen2012}
and optical properties of cavities \cite{Groblacher2009,Collette2013,Hayrynen2011c,Hayrynen2009b}.
Simple description of the quantum aspects of energy transfer is especially
interesting and challenging in lossy nanoscale systems often simulated by the
prototypical one-dimensional lossy structures.

The research of quantum optical processes in such lossy systems during the last few decades has
generated a wealth of information on the quantization of the electromagnetic (EM) field
in dielectrics and especially in layered structures.
In the context of the present work, the most relevant quantization approach
has focused on the input--output relation formalism (IORF) of the photon creation
and annihilation operators providing a detailed description of the spatial field
evolution. The formalism was originally developed for dispersionless and lossless
dielectrics by Kn\"oll \emph{et al.} \cite{Knoll1987} and soon extended for lossy and
dispersive media for the study of passive optical devices, such as dielectric plates,
interfaces, and optical cavities by several groups
\cite{Knoll1991,Allen1992,Huttner1992,Barnett1995,Matloob1995,Matloob1996}.
The complete quantization procedures studied, e.g., by Barnett \emph{et al.} \cite{Barnett1995}
fully accounting for the coupling between the EM field and the states of homogeneous lossy media clearly
highlighted that the noise and field operators in lossy systems became position
dependent. The vector potential and electric field operators in the above works
obeyed the well-known canonical commutation relation for an arbitrary choice of normal
mode functions as expected \cite{Barnett1995,Matloob1995}, but the commutation
relations of the ladder operators did not. The anomalous commutation relations
of the ladder operators were studied in several reports
\cite{Raymer2013b,Barnett1996,Ueda1994,Aiello2000,Stefano2000} but no clear resolution for the
anomalies was found apart from reaching a consensus that the anomalies were
irrelevant as long as the field commutation relations and classical field
quantities were well defined. Since then, the IORF has mainly been applied
in calculating the classical field quantities until the very recent suggestions that,
despite the early interpretations, the ladder operators and their commutation
relations might in fact relate to measurable physical properties
\cite{Collette2013}.

In this paper, we define position-dependent ladder operators associated with the vector potential
in a way that is consistent with the canonical commutation relations.
The introduced position-dependent annihilation and photon-number
operators give further insight on the local effective photon number, thermal balance, and the
formation of the local thermal equilibrium. We focus on thermal source fields, but in principle
the introduced operators also allow one to investigate fields with other kinds of quantum statistics.
We also discuss the possibility of dividing the photon number into left and right
propagating parts, and the origin of the anomalies reported in cavity operators based on this division.
The approach that is initially based on purely mathematical arguments is also shown
to result in a physically meaningful definition of the photon number that has a
very attractive and simple connection to the field temperature and thermal balance of the system.

The paper is organized as follows. In Sec.~\ref{sec:theory} the theoretical background
is briefly reviewed and the position-dependent ladder and photon-number operators 
satisfying the canonical commutation relations are defined.
To enable comparison between the conventional field operators and
the position-dependent photon-number operator we also briefly review
the concepts for the energy density and Poynting vector in
lossy, dispersive media. In addition, we show how the introduced effective
photon-number operator is related to the thermal balance of the system.
In Sec.~\ref{sec:results}, the presented theoretical concepts are applied
to study the position dependence of the photon number, electric field fluctuation,
electrical contribution of the local density of EM states,
energy density, Poynting vector, and net emission rate in the geometries
of an interface separating a lossy medium from vacuum and a vacuum cavity between lossy media.

\section{\label{sec:theory}Field quantization}
\subsection{Overview of the noise operator formalism for EM field quantization}

Field quantization in lossy dielectrics is well established, but for the purposes of
the present discussion it is convenient to briefly review some of the key results and
theoretical background for the quantization of the EM field in one dimension.
In this section, we give an overview
of the results originally presented by Matloob \emph{et al.} \cite{Matloob1995}
to lay the ground for defining the position-dependent ladder and photon-number
operators and the related discussion in the following sections.

The electromagnetic waves are considered to propagate parallel to the $x$ axis with their
transverse electric and magnetic vector operators $\hat E(x,t)$ and $\hat B(x,t)$
parallel to the $y$ and $z$ axes, respectively. The field operators are related
to the vector potential operator $\hat A(x,t)$ by the relations \cite{Matloob1995}
\begin{align}
 \hat E^+(x,\omega) & = i\omega\hat A^+(x,\omega)\label{eq:EfromA},\\
 \hat B^+(x,\omega) & = \frac{\partial}{\partial x}\hat A^+(x,\omega)\label{eq:BfromA}
\end{align}
for positive frequencies. The negative frequency parts $\hat E^-(x,\omega)$,
$\hat B^-(x,\omega)$, and $\hat A^-(x,\omega)$ are Hermitian conjugates
of the positive frequency parts.

The field operators satisfy the frequency domain Maxwell's equations and,
when the relations in Eqs.~\eqref{eq:EfromA} and \eqref{eq:BfromA} are used,
the vector potential operator can be shown to satisfy the
one-dimensional nonhomogeneous Helmholtz equation \cite{Matloob1995}
\begin{equation}
\frac{\partial^2}{\partial x^2}\hat A^+(x,\omega)+\frac{\omega^2n(x,\omega)^2}{c^2}\hat A^+(x,\omega)=-\mu_0\hat J_\mathrm{em}(x,\omega),
\label{eq:Helmholtz}
\end{equation}
where $\mu_0$ is the permeability of vacuum,
$n(x,\omega)$ is the refractive index of the medium, and
$\hat J_\mathrm{em}(x,\omega)=j_0(x,\omega)\hat f(x,\omega)$ is a Langevin noise current operator
presented in terms of the scaling factor
$j_0(x,\omega)$ and the modified Langevin force operator $\hat f(x,\omega)$, which is a
bosonic field operator defined through the following commutation relations:
\begin{align}
 [\hat f(x,\omega),\hat f^\dag(x',\omega')] & = \delta(x-x')\delta(\omega-\omega')\label{eq:commutationf1},\\
 [\hat f(x,\omega),\hat f(x',\omega')] & =[\hat f^\dag(x,\omega),\hat f^\dag(x',\omega')]=0\label{eq:commutationf2}.
\end{align}
The scaling factor of the Langevin noise current operator is given by $j_0(x,\omega)=\sqrt{4\pi\hbar\omega^2\varepsilon_0\mathrm{Im}[n(x,\omega)^2]/S}$,
where $\hbar$ is the reduced Planck's constant, $\varepsilon_0$ is the permittivity
of vacuum, and $S$ is the area of quantization in the $y$-$z$ plane \cite{Matloob1995}.
The magnitude of the scaling factor has been determined by requiring that the vector potential
and electric field operators obey the canonical equal-time commutation relation
as detailed in Ref.~\cite{Matloob1995}.

The solution to Eq.~\eqref{eq:Helmholtz} can be written in terms of the Green's function
of the Helmholtz equation as
\begin{equation}
 \hat{A}^+(x,\omega) = \mu_0\int_{-\infty}^\infty j_0(x',\omega) G(x,\omega,x')\hat f(x',\omega)dx'.
 \label{eq:solutionA}
\end{equation}
The Green's function depends on the problem geometry via the refractive index of the medium.
For example, in a homogeneous space the Green's function can be written as
\begin{equation}
 G(x,\omega,x')=\frac{ie^{ik(\omega)|x-x'|}}{2k(\omega)},
\end{equation}
where $k(\omega)=\omega n(\omega)/c$ is the wave vector.
The Green's functions for the selected layered structures are given in the Appendix.

In order to write the field operators in compact forms, we define the scaled forms
of the Green's functions:
\begin{align}
 G_\mathrm{A}(x,\omega,x') =\; & \mu_0j_0(x',\omega)G(x,\omega,x')\label{eq:GA},\\[10pt]
 G_\mathrm{E}(x,\omega,x') =\; & i\mu_0\omega j_0(x',\omega)G(x,\omega,x')\label{eq:GE},\\[6pt]
 G_\mathrm{B}(x,\omega,x') =\; & \frac{i\mu_0\omega n(x,\omega)}{c}j_0(x',\omega)\nonumber\\
                               & \times[G_\mathrm{R}(x,\omega,x')-G_\mathrm{L}(x,\omega,x')]\label{eq:GB},
\end{align}
where $G_\mathrm{E}(x,\omega,x')$ and $G_\mathrm{B}(x,\omega,x')$ are obtained from $G_\mathrm{A}(x,\omega,x')$
by using Eqs.~\eqref{eq:EfromA} and \eqref{eq:BfromA}, and 
$G_\mathrm{R}(x,\omega,x')$ and $G_\mathrm{L}(x,\omega,x')$ are the right and left
propagating parts of the Green's function identified from the factors
$e^{ikx}$ and $e^{-ikx}$,
respectively. In this paper, when treating lossless semi-infinite media,
an infinitesimal imaginary part of the refractive index is assumed in $j_0(x,\omega)$ and $G(x,\omega,x')$
and it is set to zero after calculating the integrals. Using the above definitions the
field operators are given by
\begin{align}
 \hat{A}^+(x,\omega) & = \int_{-\infty}^\infty G_\mathrm{A}(x,\omega,x')\hat f(x',\omega)dx'\label{eq:A},\\
 \hat{E}^+(x,\omega) & = \int_{-\infty}^\infty G_\mathrm{E}(x,\omega,x')\hat f(x',\omega)dx'\label{eq:E},\\
 \hat{B}^+(x,\omega) & = \int_{-\infty}^\infty G_\mathrm{B}(x,\omega,x')\hat f(x',\omega)dx'\label{eq:B}.
\end{align}
In time domain the fields are given by the inverse Fourier transforms of the frequency domain operators
as $\hat{A}(x,t)= \frac{1}{2\pi}\int_0^\infty\hat A^+(x,\omega)e^{-i\omega t}d\omega + \mathrm{H.c.}$,
where the Hermitian conjugate is the negative frequency part, and $\hat E(x,t)$ and $\hat B(x,t)$
are obtained from similar expressions.

The electric displacement field operator $\hat{D}(x,t)$ and the magnetic field strength operator $\hat{H}(x,t)$,
needed, e.g., in calculating the energy density and Poynting vector, are obtained from 
the electric field and magnetic field density operators using the constitutive relations
$\hat{D}^+(x,\omega)=\varepsilon_0\varepsilon(x,\omega)\hat{E}^+(x,\omega)$
and $\hat{B}^+(x,\omega)=\mu_0\mu(x,\omega)\hat{H}^+(x,\omega)$,
where $\varepsilon(x,\omega)$ and $\mu(x,\omega)$ are the 
position-dependent relative permittivity and permeability of the medium  \cite{Novotny2006}.

\subsection{\label{sec:operators}Ladder and photon-number operators}
In any quantum electrodynamics (QED) description, the canonical commutation relations are satisfied for
field quantities, i.e., $[\hat A(x,t),\hat E(x,t)]=-i\hbar/(\varepsilon_0S)\delta(x-x')$ \cite{Scheel1998},
but the same is not generally true for the canonical commutation relations of the ladder operators.
The dominant approach in evaluating the ladder operators has been to separate the
field operators obtained from QED either into the left and right 
propagating normal modes or into the normal modes related to the left and right inputs
and the corresponding ladder operators so that the vector potential can be written as
$\hat A^+(x,\omega)=u_\mathrm{R}(x)\hat a_\mathrm{R}(\omega) + u_\mathrm{L}(x)\hat a_\mathrm{L}(\omega)$
\cite{Barnett1996,Gruner1996,Aiello2000}.
This is tempting in view of the analogy with classical EM, but in most cases results in
ladder operators that are not unambiguously determined due to the possibility to scale
the normal modes nearly arbitrarily. More recently, also divisions accounting for
the noise contribution and more physically transparent interpretations \cite{Stefano2000}
have been reported, but none of the previously reported definitions consistently give
the canonical commutation relations for the ladder operators.

We adopt a different starting point that ensures the preservation of the canonical
commutation relations by simply writing
\begin{equation}
 \hat A^+(x,\omega)=C(x,\omega)\hat a(x,\omega),
 \label{eq:initial}
\end{equation}
where $\hat a(x,\omega)$ is the position dependent photon annihilation operator and
$C(x,\omega)$ is a normalization factor that corresponds to the classical mode function
defined simultaneously for all the source points.
Solving Eq.~\eqref{eq:initial} for $\hat a(x,\omega)$ and using Eq.~\eqref{eq:A}
for $\hat A(x,\omega)$ gives
\begin{equation}
 \hat a(x,\omega)=\frac{\int G_\mathrm{A}(x,\omega,x')\hat f(x',\omega)dx'}{C(x,\omega)}.
 \label{eq:totalfielda}
\end{equation}
The canonical commutation relation for the ladder operators must fulfill
$[\hat a(x,\omega),\hat a^\dag(x,\omega')]=\delta(\omega-\omega')$.
Substituting the field annihilation operator in Eq.~\eqref{eq:totalfielda}
to the canonical commutation relation gives
\begin{equation}
 [\hat a(x,\omega),\hat a^\dag(x,\omega')]=\frac{\int |G_\mathrm{A}(x,\omega,x')|^2dx'}{|C(x,\omega)|^2}\delta(\omega-\omega'),
 \label{eq:acommutation}
\end{equation}
which can be used to determine $C(x,\omega)$ so that the canonical commutation
relation holds at any position. The phase factor does not play a role in our
calculations so, choosing $C(x,\omega)$ to be real and positive, we have
\begin{equation}
 C(x,\omega)=\sqrt{\int |G_\mathrm{A}(x,\omega,x')|^2dx'}.
\end{equation}
This can be expressed in terms of the imaginary part of the Green's function as \cite{Dung1998,Stefano2000}
\begin{equation}
 C(x,\omega)=\sqrt{\frac{4\pi\hbar}{\varepsilon_0c^2S}\mathrm{Im}[G(x,\omega,x)]}.
\end{equation}
Since the electric field operator is related to the vector potential
operator by Eq.~\eqref{eq:EfromA}, the electric field
operator is $\hat E^+(x,\omega)=i\omega C(x,\omega)\hat a(x,\omega)$.
The time domain electric field operator is naturally given by
taking the inverse Fourier transform of the frequency domain operator as
\begin{equation}
 \hat E(x,t) = \frac{i}{2\pi}\int_0^\infty\!\!\! \omega C(x,\omega)\Big[\hat a(x,\omega)e^{-i\omega t}-\hat a^\dag(x,\omega)e^{i\omega t}\Big]d\omega.
 \label{eq:Ea}
\end{equation}

The photon number operator is given in
terms of the ladder operators as
$\hat n(x,\omega)=\int\hat a^\dag(x,\omega)\hat a(x,\omega')d\omega'$
and its expectation value is expressed in terms of the Green's function as
\begin{equation}
 \langle\hat n(x,\omega)\rangle=\frac{\int |G_\mathrm{A}(x,\omega,x')|^2\langle\hat\eta(x',\omega)\rangle dx'}{|C(x,\omega)|^2}.
 \label{eq:photonnumber}
\end{equation}
Here we have defined a source field photon-number operator as
$\hat \eta(x,\omega) =\linebreak \int\hat f^\dag(x,\omega)\hat f(x',\omega')\,dx'd\omega'$
and assumed that the noise operators at different positions and at different frequencies
are uncorrelated so that the source field photon-number expectation value at
position $x$ of a thermally excited medium is
\begin{equation}
 \langle\hat \eta(x,\omega)\rangle = \frac{1}{e^{\hbar\omega/(k_\mathrm{B}T(x))}-1},
 \label{eq:sourcefieldn}
\end{equation}
where $k_\mathrm{B}$ is the Boltzmann constant and
$T(x)$ is the position-dependent temperature of the medium.
In the case of thermal fields the photon-number operator in Eq.~\eqref{eq:photonnumber}
also allows one to calculate an effective local field temperature for the electric field as
$T(x,\omega)=\hbar\omega/\{k_\mathrm{B}\ln[1+1/\langle\hat n(x,\omega)\rangle]\}$.
As discussed later in Sec.~\ref{sec:results}, the local photon number and the associated
field temperature should be experimentally observable with a suitable measurement setup.

The above definition of the photon annihilation operator
may seem purely mathematical, but it will be shown to have
clear physical implications. In particular, as the ladder
operators are essentially defined to  be proportional to the
vector potential (and therefore also the electric field),
the photon-number operator can be considered as an effective
operator that partly reflects the physical properties of the
electric field. For instance, under certain nonequilibrium
conditions studied in Sec.~\ref{sec:results}, the photon number
can oscillate due to the interference seen in the electric
fields. It is also found that without the above exact form
for the position-dependent normalization coefficient
the resulting photon number expectation value
oscillates near material interfaces at thermal equilibrium. The properly
normalized annihilation operator defined in Eqs.~\eqref{eq:totalfielda} and
\eqref{eq:acommutation} always results in a photon-number
expectation value that is constant everywhere
at thermal equilibrium.

\subsection{Fields emitted by the left and right source domains}
For mathematical comparison of the proposed position-dependent photon annihilation
operator with previously used IORF approaches,
we separate the annihilation operator in Eq.~\eqref{eq:totalfielda} into
parts $\hat a_\mathrm{R}(x,\omega)$ and $\hat a_\mathrm{L}(x,\omega)$ arising from the emission
from the half spaces $]\!-\infty,x]$ and $[x,\infty[$, respectively. In the reflectionless
case these operators also correspond to the right and left propagating modes.
The left and right source domain operators are given by
\begin{align}
 \hat a_\mathrm{R}(x,\omega) & =\frac{\int_{-\infty}^x G_\mathrm{A}(x,\omega,x')\hat f(x',\omega)dx'}{C(x,\omega)}\label{eq:ar},\\
 \hat a_\mathrm{L}(x,\omega) & =\frac{\int_x^\infty G_\mathrm{A}(x,\omega,x')\hat f(x',\omega)dx'}{C(x,\omega)}\label{eq:al}.
\end{align}
These operators account for the normalized contribution of each source
point located to the left and right from the point of observation $x$ but generally
contain both right and left propagating terms. Alternatively to Eqs.~\eqref{eq:ar} and \eqref{eq:al}
one could also separate $\hat a(x,\omega)$ to the conventional left
and right propagating parts by separately accounting for the left and right propagating parts
of the Green's function $G_\mathrm{L}(x,\omega,x')$ and $G_\mathrm{R}(x,\omega,x')$,
introduced in Eq.~\eqref{eq:GB}.
However, this would not result in the canonical commutation
relations except for some special cases.
In the case of operators in Eqs.~\eqref{eq:ar} and \eqref{eq:al},
the commutation relations are generally given by
\begin{align}
 &[\hat a_\mathrm{R}(x,\omega),\hat a_\mathrm{R}^\dag(x',\omega')]\nonumber\\
 &\hspace{0.5cm} = \frac{\int_{-\infty}^{\min(x,x')} G_\mathrm{A}(x,\omega,y)G_\mathrm{A}^*(x',\omega',y)dy}{C(x,\omega)C(x',\omega')}\delta(\omega-\omega'),\\
 &[\hat a_\mathrm{L}(x,\omega),\hat a_\mathrm{L}^\dag(x',\omega')]\nonumber\\
 &\hspace{0.5cm} = \frac{\int_{\max(x,x')}^\infty G_\mathrm{A}(x,\omega,y)G_\mathrm{A}^*(x',\omega',y)dy}{C(x,\omega)C(x',\omega')}\delta(\omega-\omega'),\\
 &[\hat a_\mathrm{R}(x,\omega),\hat a_\mathrm{L}^\dag(x',\omega')]\nonumber\\
 &\hspace{0.5cm} = [\hat a_\mathrm{L}(x',\omega'),\hat a_\mathrm{R}^\dag(x,\omega)]^*\nonumber\\
 &\hspace{0.5cm} = \frac{\theta(x-x')\int_{x'}^x G_\mathrm{A}(x,\omega,y)G_\mathrm{A}^*(x',\omega',y)dy}{C(x,\omega)C(x',\omega')}\delta(\omega-\omega')\label{eq:lastcommutator},
\end{align}
where the integrals are taken over the source domain region that is common
for both operators. In regions where the field only propagates in one direction
[i.e., $G(x,\omega,x')$ only contains terms of the form
$e^{ikx}$ or $e^{-ikx}$], the canonical commutation relation
arises naturally and allows straightforwardly separating the photon annihilation
operator into the left and right propagating parts. If the
Green's function contains both $e^{ikx}$ and $e^{-ikx}$,
the photon annihilation operators $\hat a_\mathrm{R}(x,\omega)$
and $\hat a_\mathrm{L}(x,\omega)$ do not commute, and thus,
defining the left and right source domain operators in Eqs.~\eqref{eq:ar} and \eqref{eq:al}
while satisfying canonical commutation relations is not possible.
The same applies to defining the corresponding left and right
propagating operators. This is fundamentally the origin
of the anomalies found in the commutation relations of
the photon annihilation operator \cite{Barnett1996,Ueda1994,Aiello2000},
and also the origin of the photon-number oscillations observed in Sec. \ref{sec:results}.
Note, however, that the annihilation operators of the left and right source domains
commute at a common position since the integration domain in the commutation
relation in Eq.~\eqref{eq:lastcommutator} is of zero length when $x=x'$.
Therefore, the photon-number operator can always be separated into
parts corresponding to the left and right source domains.

\subsection{Poynting vector and energy density}
For additional physical insight, we will compare the position dependence of the photon-number
expectation value following from Eq.~\eqref{eq:photonnumber}
to the well-known fluctuations in the electric field, energy fluxes, and energy densities.
The one-dimensional quantum optical Poynting vector operator
is defined in terms of the positive and negative frequency parts of the electric 
and magnetic field operators as $\hat{S}(x,t)=\hat{E}^-(x,t)\hat{H}^+(x,t)+\hat{H}^-(x,t)\hat{E}^+(x,t)$
\cite{Janowicz2003,Loudon2000}.
Substituting the positive and negative frequency parts of the given electric and magnetic field
operators and calculating the expectation value gives the spectral component of the
quantum optical Poynting vector as
\begin{equation}
 \Scale[0.95]{\displaystyle\langle\hat{S}(x,t)\rangle_\omega =\frac{1}{2\pi^2}\!\int\!\mathrm{Re}[G_\mathrm{E}^*(x,\omega,x')G_\mathrm{H}(x,\omega,x')]\langle\hat\eta(x',\omega)\rangle dx',}
 \label{eq:poynting}
\end{equation}
where $G_\mathrm{H}(x,\omega,x')=G_\mathrm{B}(x,\omega,x')/\boldsymbol{(}\mu_0\mu(x,\omega)\boldsymbol{)}$.
The brackets denote the expectation value over all states
resulting in the source field photon-number expectation value of Eq.~\eqref{eq:sourcefieldn},
and the subscript $\omega$ denotes the spectral component of the Poynting vector,
i.e., the integrand when the total Poynting vector is expressed as an integral
over positive frequencies.

Defining the electromagnetic energy density in a general, lossy, dispersive
medium has been challenging and common definitions have, e.g., led
to negative values for the energy density \cite{Ziolkowski2001}
or energy density expressions that are valid only for
Lorentz dielectrics \cite{Ruppin2002}.
We use a recently generalized definition of the average
total energy density introduced by Vorobyev \cite{Vorobyev2012,Vorobyev2013}.
However, as the definition was originally developed for time averages
assuming a classical sinusoidal electromagnetic field, we
use a prefactor $1/2$ instead of $1/4$ for general non-time-averaged
fields giving the energy density as
\begin{align}
 \langle\hat u(x,t)\rangle_\omega =\;&\frac{\varepsilon_0}{2}\Big(1+\Big|\frac{\partial(\chi_e(x,\omega)\omega)}{\partial\omega}\Big|\Big)\langle\hat{E}(x,t)^2\rangle_\omega\nonumber\\[4pt]
 &+\frac{\mu_0}{2}\Big(1+\Big|\frac{\partial(\chi_m(x,\omega)\omega)}{\partial\omega}\Big|\Big)\langle\hat{H}(x,t)^2\rangle_\omega.
 \label{eq:edensity}
\end{align}
Here $\chi_e(x,\omega)=\varepsilon(x,\omega)-1$ and $\chi_m(x,\omega)=\mu(x,\omega)-1$
are the position-dependent electric and magnetic susceptibilities.

The spectral components of the electric and magnetic field fluctuations in Eq.~\eqref{eq:edensity} can
be directly calculated by using the conventional Green's function expressions for
the field operators giving
\begin{equation}
 \langle\hat{E}(x,t)^2\rangle_\omega=\frac{1}{2\pi^2}\int|G_\mathrm{E}(x,\omega,x')|^2\Big(\langle\hat\eta(x',\omega)\rangle+\frac{1}{2}\Big)dx'\label{eq:efluct}
\end{equation}
for the electric field. The expression for the magnetic field strength fluctuation is obtained
with $G_\mathrm{E}(x,\omega,x')\rightarrow G_\mathrm{H}(x,\omega,x')$.
It is also possible to express the electric field fluctuation directly in terms
of the introduced position-dependent photon-number operator as
\begin{equation}
 \langle\hat E(x,t)^2\rangle_\omega=\frac{\hbar\omega}{\varepsilon_0}\rho(x,\omega)\Big(\langle\hat n(x,\omega)\rangle+\frac{1}{2}\Big),
\end{equation}
where we have used the conventional definition for the electrical contribution of the
local density of EM states (electric LDOS) as \cite{Joulain2003}
\begin{equation}
 \rho(x,\omega)=\frac{\varepsilon_0\omega}{2\pi^2\hbar}|C(x,\omega)|^2=\frac{2\omega}{\pi c^2S}\mathrm{Im}[G(x,\omega,x)].
 \label{eq:ldos}
\end{equation}

\subsection{\label{sec:balance}Thermal balance}
For a better understanding of the introduced photon-number concept it is essential to consider
its contribution to local energy balance at position $x$. The Poynting theorem
relates the local power dissipation and generation
to the current density and electric field at position $x$ and to
the divergence of the Poynting vector \cite{Jackson1999}. Therefore, in one dimension, the spectral
energy transfer rate $\langle Q(x,t)\rangle_\omega$, i.e., the spectral net emission between
the field and the local medium, is given by
\begin{equation}
 \langle Q(x,t)\rangle_\omega=\frac{\partial}{\partial x}\langle\hat S(x,t)\rangle_\omega=-\langle\hat J(x,t)\hat E(x,t)\rangle_\omega.
 \label{eq:conservation}
\end{equation}
The current density term $\hat J(x,t)=\hat J_\mathrm{abs}(x,t)-\hat J_\mathrm{em}(x,t)$
consists of parts corresponding to absorption and emission.
The photon emission is described by the Langevin noise current operator
$\hat J_\mathrm{em}(x,\omega)=j_0(x,\omega)\hat f(x,\omega)$
introduced as a field source in Eq.~\eqref{eq:Helmholtz}.
The photon absorption term is of the form
$\hat J_\mathrm{abs}(x,\omega)=\varepsilon_0\omega\,\mathrm{Im}[n(x,\omega)^2]\hat E^+(x,\omega)$
so that the absorption rate is proportional to the square of the electric field,
or equivalently the electric field fluctuation.
Substituting the time domain current terms (i.e., the Fourier
transforms of the frequency domain terms) to Eq.~\eqref{eq:conservation} and calculating
the expectation values over source field photon-number states gives the
local net emission rate in terms of the photon numbers of
the source and the total electromagnetic fields as
\begin{equation}
 \Scale[0.96]{\displaystyle\langle Q(x,t)\rangle_\omega=\hbar\omega^2\mathrm{Im}[n(x,\omega)^2]\rho(x,\omega)[\langle\hat\eta(x,\omega)\rangle-\langle\hat n(x,\omega)\rangle].}
 \label{eq:divP}
\end{equation}
Equation \eqref{eq:divP} directly shows that the local net emission rate is
zero only if the material is lossless $(\mathrm{Im}[n(x,\omega)^2]=0)$, the electric LDOS is zero $[\rho(x,\omega)=0]$,
or the field is in local thermal equilibrium $[\langle\hat n(x,\omega)\rangle=\langle\hat\eta(x,\omega)\rangle]$.
Equation \eqref{eq:divP} also nicely separates the effect of temperature and
wave features in the local net emission rate: The effect of temperature is described
by the photon-number operators and the effect of wave features is described by the
imaginary part of the Green's function.
In addition, Eq.~\eqref{eq:divP} is essentially the equivalent of the multiprobe
Landauer-B\"uttiker formula \cite{Buttiker1992,Saaskilahti2013} generalized for photons and continuous media.
In resonant systems where the energy exchange is dominated by a narrow frequency band,
condition $\langle\hat Q(x,\omega)\rangle_\omega=0$ can be used
to  approximately determine the steady-state temperature of a weakly interacting resonant
particle \cite{Bohren1998}. This leads to concluding that in order to reach a thermal
balance with the field, the particle must reach a temperature
that is equal to the effective field temperature so that the term
$\langle\hat\eta(x,\omega)\rangle-\langle\hat n(x,\omega)\rangle$ disappears.

\section{\label{sec:results}Results}

\begin{figure*}
\includegraphics[width=\textwidth]{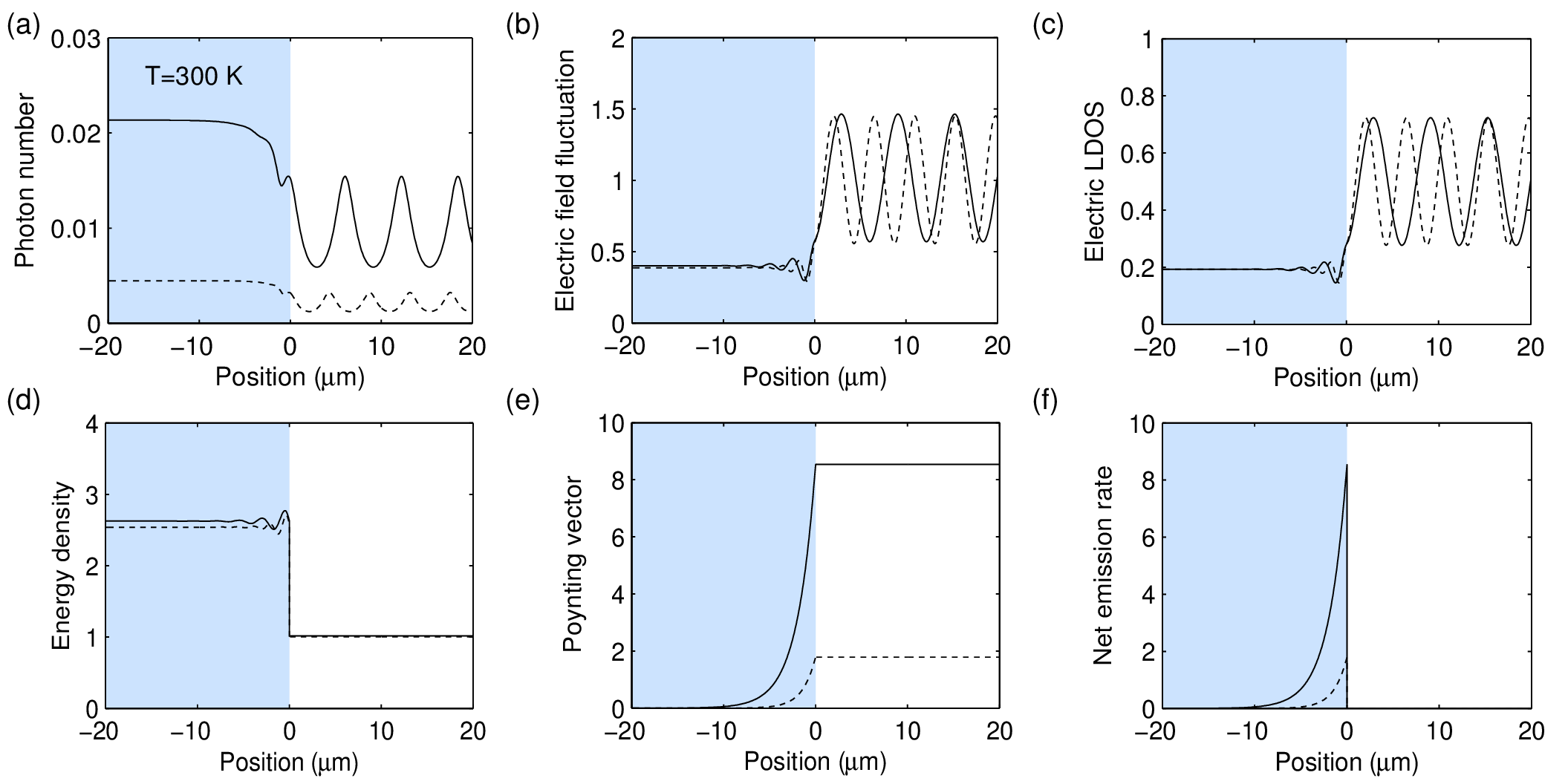}
\caption{\label{fig:semi}(Color online) (a) The effective photon number, (b) electric field fluctuation, (c) electric LDOS, (d) energy density,
(e) Poynting vector, and (f) net emission rate in the vicinity of an interface separating a lossy medium with refractive index
$n_1=2.5+0.5i$ and temperature 300 K from vacuum at 0 K.
The solid lines correspond to the photon energy $\hbar\omega=0.10$ eV ($\lambda=12.4$ $\mu$m)
and the dashed lines correspond to the photon energy $\hbar\omega=0.14$ eV ($\lambda=8.86$ $\mu$m).
The electric field fluctuation is given in the
units of $\hbar\omega/(2\pi\varepsilon_0cS)$,
the electric LDOS in the units of $2/(\pi cS)$,
the energy density in the units of $\hbar\omega/(2\pi cS)$,
the Poynting vector in the units of $\hbar\omega/(1000\pi S)$,
and the net emission rate in the units of $\hbar\omega^2/(1000\pi cS)$.}
\end{figure*}

To investigate the physical implications of the concepts presented in Sec.~\ref{sec:theory}
we compare the position-dependent photon number to the corresponding electric field fluctuation, electric
LDOS, energy density, Poynting vector, and net emission rate in two geometries: a single interface
separating a lossy medium from vacuum and a vacuum cavity formed between two
lossy media. The definitions
for the electric field fluctuation, energy density, electric LDOS, and Poynting vector coincide
with the conventional Green's function based definitions. Instead, the position-dependent
photon-number operator predicts different results.

\subsection{Dielectric-vacuum interface}

The single interface structure consists of a semi-infinite lossy
medium with refractive index $n=2.5+0.5i$ at temperature $T_1=300$ K. The field
incident from vacuum is a vacuum field with $T_2=0$ K.
Figure \ref{fig:semi} shows the effective photon number, electric field fluctuation, electric LDOS,
energy density, Poynting vector, and net emission rate as a function of position for the single
interface geometry for photon energies $\hbar\omega=0.10$ eV and
$\hbar\omega=0.14$ eV, corresponding to wavelengths $\lambda=12.4$ $\mu$m and $\lambda=8.86$ $\mu$m.
The expectation value of the photon number operator in Eq.~\eqref{eq:photonnumber}
is plotted in Fig.~\ref{fig:semi}(a). The effective photon number has a strong position dependence
and it oscillates both in the vacuum and inside the lossy medium. In the lossy medium
the oscillations are damped and the photon number saturates to a constant
value far from the interface. The damping takes place over a distance of
10 $\mu$m that approximately corresponds to the penetration depth of radiation.
The oscillations in the vacuum can be explained by separately considering
the interference of the field component terms generated by the lossy half-space
and the incident field from the vacuum, i.e., the left and right source domain contributions
$x'\in]\!-\infty,0]$ and $x'\in[0,\infty[$ in Eq.~\eqref{eq:photonnumber}.
On the vacuum side, the term $|G(x,\omega,x')|^2$ in Eq.~\eqref{eq:photonnumber}
is constant for the right source domain contribution as it describes
a simple propagating wave in that region. The contribution from the left
source domain involves a Green's function with standing wave features
and therefore $|G(x,\omega,x')|^2$ for the left source domain oscillates.
To preserve the canonical commutation relations,
however, the oscillations are compensated by changes in the contribution
of the fields generated by the lossy medium. This is a direct
mathematical consequence imposed by the preservation of the canonical
commutation relations and results in variations of the contributions
of the vacuum field at $T_2=0$ K and the lossy layer at $T_1=300$ K
as a function of position even in the vacuum. In lossy media, similar
oscillations occur but they are quickly damped by the losses.
The oscillations are expected to be related to defining the ladder operators so 
that they are proportional to the vector potential and the electric field. This implies 
that the resulting photon number is in fact a quantity that mainly reflects the features 
related to electric field and field--matter interactions involving electrical dipoles.

The corresponding position-dependent magnitude of the electric field
fluctuation in Eq.~\eqref{eq:efluct} is plotted in Fig.~\ref{fig:semi}(b). The magnitude
in the vacuum is larger than inside the lossy medium. This is not
always the case, since it depends on the refractive index of the medium and
the source field temperature. The field fluctuations also exhibit
similar oscillations as the photon numbers in Fig.~\ref{fig:semi}(a)
but the minima of the electric field fluctuations coincide with
the maxima of the photon-number oscillations. In the lossy layer the
fluctuations are quickly damped to a constant value. Overall,
oscillations in the photon number are accompanied by the oscillations
in the electric field fluctuations, but there is no direct correspondence
between the two.

The electric LDOS in Eq.~\eqref{eq:ldos} is plotted in Fig. \ref{fig:semi}(c).
The electric LDOS graphs clearly resemble the electric field fluctuations in Fig. \ref{fig:semi}(b)
oscillating in the vacuum and saturating to constant values inside the lossy medium.
In contrast to the field fluctutations, the electric LDOS is temperature independent and not directly
proportional to the fluctuations. However, since the zero-point-field contribution dominates at
chosen temperatures and frequencies, the oscillating photon number has only a small effect
on the electric field fluctuations. The oscillations in the electric LDOS describe the modified
emission rate due to interference better known as 
the Purcell effect in cavities.

The energy density in Eq.~\eqref{eq:edensity} is shown in Fig.~\ref{fig:semi}(d).
Inside the lossy medium, the energy density exhibits damped oscillations and
in the vacuum it is perfectly constant since the electric and magnetic field
contributions in the energy density are equal and out of phase so that the total
energy density becomes constant. In the lossy medium the electric field contribution
dominates in the energy density. This leads to oscillations that
resemble the forms seen in the electric field fluctuation graphs in
Fig.~\ref{fig:semi}(b).

The Poynting vector in Eq.~\eqref{eq:poynting} is shown in Fig.~\ref{fig:semi}(e).
No net power propagates deep inside the lossy material since the
absorption and emission of thermal photons is in balance, and therefore,
the Poynting vector equals zero. The field generated in the material near
the interface can propagate into the vacuum so that the Poynting vector
grows exponentially and preserves its constant value through the vacuum
indicating a constant radiative heat flow out of the lossy half-space.

The net emission rate given by Eq.~\eqref{eq:divP} is plotted
in Fig.~\ref{fig:semi}(f). The net emission is naturally zero in the vacuum.
The positive net emission in the lossy medium
near the interface denotes that the rate of photon emission
outweighs the rate of photon absorption. Inside the lossy medium
the net emission rate decays to zero since the photon emission
and absorption become balanced. However, no oscillations appear
in the net emission rate. This is due to the infinitely large uniform
half spaces and thermal field statistics. In the presence 
of a nonuniform temperature profile in the lossy medium
(not shown) the oscillations seen in
the electric LDOS in Fig.~\ref{fig:semi}(c) are also present in the
net emission rate in the lossy medium.

\begin{figure*}
\includegraphics[width=\textwidth]{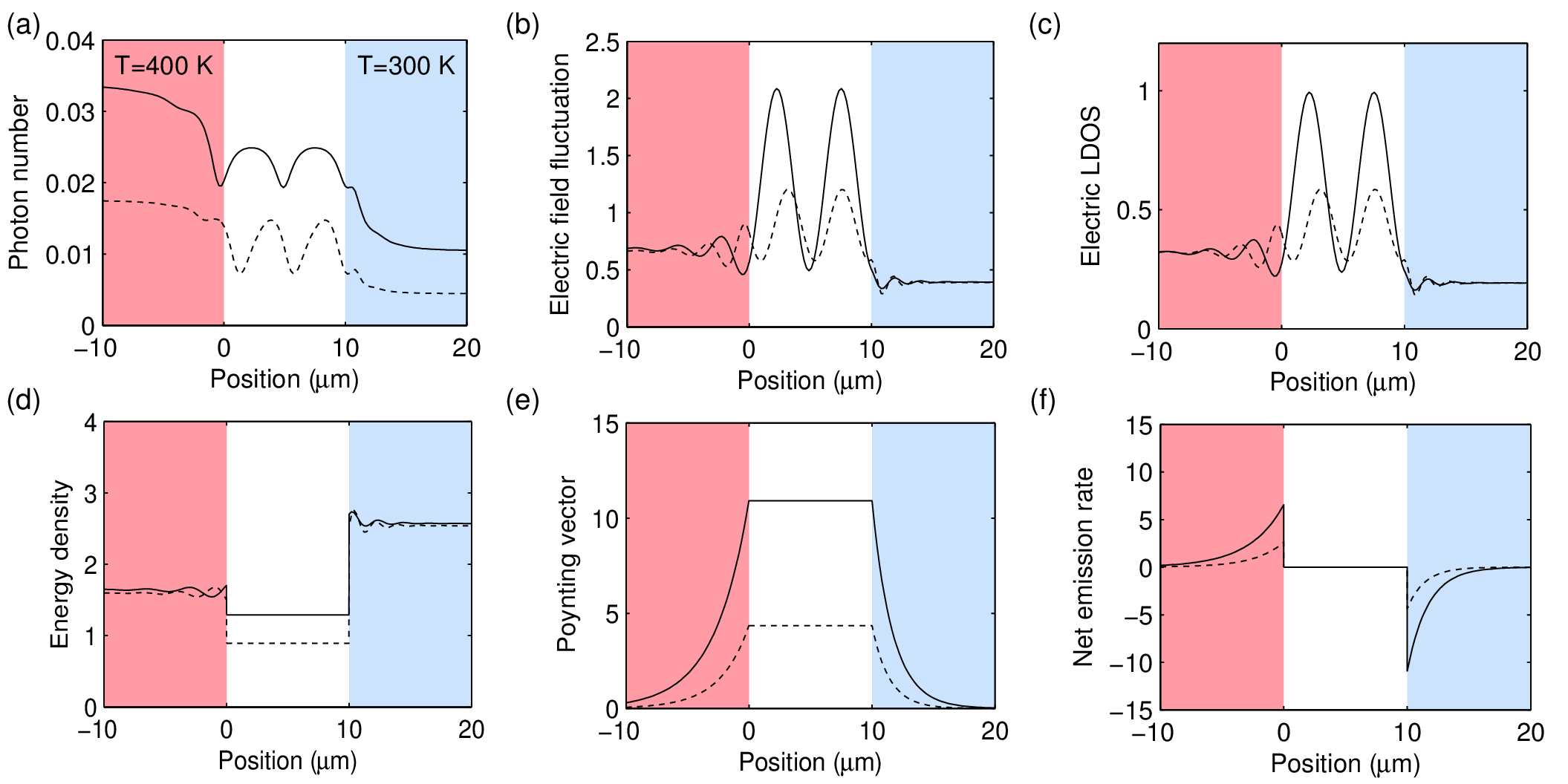}
\caption{\label{fig:slab}(Color online) (a) The effective photon number, (b) electric field fluctuation,
(c) electric LDOS, (d) energy density, (e) Poynting vector, and (f) net emission rate
in the vicinity of a vacuum gap separating lossy media with refractive indices
$n_1=1.5+0.3i$ and $n_2=2.5+0.5i$ at temperatures 400 and 300 K. The width of the cavity is 10 $\mu$m.
The solid lines correspond to the second resonant photon energy $\hbar\omega=0.118$ eV ($\lambda=10.5$ $\mu$m)
and the dashed lines correspond to the off-resonant energy $\hbar\omega=0.140$ eV ($\lambda=8.86$ $\mu$m).
The electric field fluctuation is given in the units of
$\hbar\omega/(2\pi\varepsilon_0cS)$, the electric LDOS in the units of $2/(\pi cS)$,
the energy density in the units of $\hbar\omega/(2\pi cS)$,
the Poynting vector in the units of $\hbar\omega/(1000\pi S)$,
and the net emission rate in the units of $\hbar\omega^2/(1000\pi cS)$.}
\end{figure*}

\subsection{Vacuum gap between lossy media}

As a slightly more complex example we also investigate a cavity structure which consists of two semi-infinite media
with refractive indices $n_1=1.5+0.3i$ and $n_2=2.5+0.5i$ at temperatures
$T_1=400$ K and $T_2=300$ K, separated by a vacuum gap of thickness $10$ $\mu$m.
Figure \ref{fig:slab} shows the effective photon number, electric field fluctuation, electric LDOS,
energy density, Poynting vector, and net emission rate as a function of position
for the cavity geometry for photon energies $\hbar\omega=0.118$ eV ($\lambda=10.5$ $\mu$m)
corresponding to the second resonance of the cavity and $\hbar\omega=0.140$ eV ($\lambda=8.86$ $\mu$m)
that is off-resonant. As in the case of results for the single interface geometry
presented above, there are periodic oscillations in the effective photon number as presented
in Fig.~\ref{fig:slab}(a). For the second resonant energy the photon number has two
peaks in the cavity as expected. At resonant energies observing oscillations in the
photon number, however, requires an asymmetric cavity: If the refractive indices
of the left and right lossy media were equal (not shown), the photon-number
oscillations inside the cavity would completely disappear for
resonant frequencies. On the left and right of the cavity in Fig.~\ref{fig:slab}(a),
the photon number again saturates to constant values depending
on the source field temperature.

The corresponding electric field fluctuations are presented in Fig.~\ref{fig:slab}(b).
As in the case of the single interface geometry, the magnitudes of the
electric field fluctuations are larger in the vacuum than in the lossy medium.
The extrema of the electric field fluctuations are located at the same
positions as the extrema of the photon number in Fig.~\ref{fig:slab}(a).
Inside the lossy media, the oscillations are again damped and eventually
reach constant values, which are different
on the left and right due to the different source field temperatures.

The electric LDOS is plotted in Fig. \ref{fig:slab}(c).
Again, the electric LDOS oscillates in the vacuum and saturates to
constant values in the lossy media resembling the electric field fluctuation
in Fig.~\ref{fig:slab}(b). However, there are some differences.
For example, the difference in the oscillation magnitudes
of the resonant and off-resonant energies in the electric LDOS in the vacuum
is larger than the corresponding difference in the electric field fluctuation
in Fig.~\ref{fig:slab}(b). This is due to the effect of the oscillating photon
number in the electric field fluctuation. The effect is hardly visible
in the electric field fluctuation in Fig.~\ref{fig:slab}(b), but becomes larger
when the temperature of the left medium is increased (not shown).

The energy density shown in Fig.~\ref{fig:slab}(d) exhibits damped
oscillations inside the lossy medium and in the vacuum it is perfectly
constant as in the case of the single interface geometry.
In the lossy media the energy density due to the
electric field contribution is again larger than in the vacuum
and leads to oscillations that
resemble the form seen in the electric field fluctuation graph in
Fig.~\ref{fig:slab}(b). Note that the energy density is
smaller in the left medium than in the right medium even though the left
medium has a higher temperature. This is due to the
effect of electric susceptibility which is larger
at the right medium. However, if the temperature is increased on
the left the energy density on the left at some point naturally
exceeds the energy density on the right.

The Poynting vector is plotted in Fig.~\ref{fig:slab}(e). In the vacuum gap,
the Poynting vector is again constant since there is no dissipation.
The positivity of the Poynting vector denotes
net energy transfer towards the medium at lower temperature.
Inside the lossy media, the Poynting vector
asymptotically reaches zero far from the interfaces.
The damping of the Poynting vector is faster in the right medium than in the left medium due
to the larger imaginary part of the refractive index.

The net emission rate is shown in Fig.~\ref{fig:slab}(f).
Again the net emission rate is zero in the vacuum and it
decays to zero inside the lossy media. Positive (negative)
values denote that the rate of photon emission (absorption)
outweighs the rate of photon absorption (emission).
The higher absolute value of the net emission rate in
the lossy medium with higher imaginary part of the refractive
index is related to larger loss and the faster damping of the Poynting vector
in Fig.~\ref{fig:slab}(e). In the presence of small losses inside
the cavity (not shown) the net emission rate exhibits
oscillations similar to the oscillations of the electric
LDOS in Fig.~\ref{fig:slab}(c). This directly reflects the Purcell effect
and position-dependent emission rate of particles placed in the cavity.

The proposed position-dependent ladder and photon-number operators predict
that the effective photon numbers oscillate with respect to the positions as shown in
Figs.~\ref{fig:semi}(a) and \ref{fig:slab}(a). 
In contrast to the field quantities, the effective photon number provides
a simple metric for finding the thermal balance formed due to interactions
taking place through the electric field and electric dipoles as detailed in
Eq.~\eqref{eq:divP}. Since the photon number
as defined in this work is also expected to be directly related both
to local temperature and rate of energy exchange taking place between the
electric field and the dipoles constituting the lossy materials, we 
expect that the predicted photon-number oscillations can also be measured.
A good candidate for such a measurement would be a setup consisting of a
high-quality-factor cavity where
the magnitude of oscillations as well as the incoming radiation
on the left and right can be better controlled than in the single interface case.
Photon-number measurements have already been performed in cavities \cite{Maitre1997},
but not in nonequilibrium conditions which is necessary for the oscillations of the effective
photon number with respect to the position. If the cavity is asymmetric,
the photon-number oscillations are more easily observable
than in a symmetric cavity where the photon-number oscillations
can disappear at resonant frequencies. If one
had a thermometer that resonantly couples to one of the cavity resonances and
is anisotropic so that it is capable of detecting only the single perpendicular mode
studied in this work, then the predicted photon-number oscillations could be easily
detected experimentally. In reality the measurement configuration is naturally
more complex as controlling the directivity of a nanoscale 
thermometer may not be that straightforward, but similar
effects are nevertheless also expected to be observable in fully three-dimensional
configurations. As the taken approach is very general, generalizing the model
to three dimensions is expected to be straightforward.
In this paper we have investigated only thermal source fields in detail, but
the introduced operators are expected to enable also a much more general description
of the quantized fields obeying other kinds of quantum statistics.
For example, replacing the noise operators with operators describing partly
saturated emitters \cite{Hayrynen2010b} could result in a simple but realistic
quantum description of the generation of laser fields, and furthermore,
the use of nonlinear source field operators \cite{Hayrynen2012b,Hayrynen2010c}
could even allow modeling single photon sources and detectors.

\section{\label{sec:conclusions}Conclusions}
We have defined and studied position-dependent ladder operators for the electromagnetic
field in a way that is consistent with the canonical commutation relations
and also enables a physically meaningful definition of an effective position-dependent
photon-number operator that has a very attractive and simple connection to the
electric field, the temperature, and thermal balance of the system.
The introduced operators predict oscillations in the expectation value of the effective
photon number even in simple one-dimensional
layered geometries. Since the effective photon number is expected to be directly related to
experimentally observable local temperatures in measurements in which the field-matter
interaction is dominated by the coupling to the electric field, this suggests that the
oscillations could be experimentally observable by measuring similar
oscillations in the temperature. This essentially differentiates our definition of the
effective photon number from the conventional position-independent definitions that
do not take the physical properties of the electric field similarly into account.

The introduced ladder and photon-number operators provide an additional means
to quantify the electric field and give further insight into the local
photon-number balance associated with the electric field and the formation
of the local thermal equilibrium.
The additional insight provided by the approach also allows improved understanding
of the challenges and physical implications of separating the photon annihilation operator
into parts emitted by the left and right source domains or left and right propagating parts as done in
many seminal quantization schemes. If the requirement of canonical commutation
relations and ease of physical interpretation are not relaxed, 
the separation is not generally possible. This is in agreement with the
anomalies found in the commutation relations of
the photon annihilation operator. However, possibly the greatest opportunities enabled by the
new formulation and the consistent definitions of the ladder and photon-number
operators are in extending the description to quantum systems
that are not limited to thermal fields. This could, e.g., enable a more versatile description
of the noise and measurement backaction in complex and realistic quantum systems.

\begin{acknowledgments}
This work has in part been funded by the Academy of Finland and the Aalto Energy Efficiency Research Programme.
\end{acknowledgments}

\appendix*

\section{Green's functions}
\subsection{Single interface}
The single interface geometry consists of media with refractive indices
$n_1$ ($x<0$) and $n_2$ ($x>0$). The corresponding wave vectors are $k_1$
and $k_2$. The Green's function for the single interface geometry
is given in the left half-space ($x<0$) by \cite{Matloob1996}
\begin{align}
G_{\{x<0\}}(x,\omega,x')=\;&\frac{i\,\theta(-x')}{2k_1}\Big( e^{ik_1|x-x'|}+re^{-ik_1(x+x')} \Big)\nonumber\\
  &+\frac{i\,\theta(x')}{2k_2}t'e^{-i(k_1x-k_2x')}
 \label{eq:greensemi1}
\end{align}
and on the right half-space ($x>0$) by
\begin{align}
G_{\{x>0\}}(x,\omega,x')=\;&\frac{i\,\theta(-x')}{2k_1}te^{i(k_2x-k_1x')}\nonumber\\
  &+\frac{i\,\theta(x')}{2k_2}\Big(e^{ik_2|x-x'|}+r'e^{ik_2(x+x')} \Big),
 \label{eq:greensemi2}
\end{align}
where $\theta(x)$ is the step function and $r$ and $t$ are the Fresnel coefficients for
reflection and transmission on the left given for normal incidence by
\begin{equation}
 r=\frac{n_1-n_2}{n_1+n_2},\hspace{0.5cm}
 t=\frac{2n_1}{n_1+n_2}.
 \label{eq:fresnel}
\end{equation}
The reflection and transmission coefficients on the right $r'$ and $t'$ are
obtained by switching the indices 1 and 2 in the expressions of $r$ and $t$.

In the Green's functions of both half-spaces in Eqs.~\eqref{eq:greensemi1} and \eqref{eq:greensemi2},
the first term describes the field generated by the source point in
the left half-space and the second term describes the field generated by the
source point in the right half-space.

\subsection{Two interfaces}
The slab geometry consists of media with refractive indices
$n_1$ ($x<0$), $n_2$ ($0<x<d$), and $n_3$ ($x>d$). The corresponding wave vectors are
$k_1$, $k_2$, and $k_3$. The Green's function for the slab geometry
is given in the left medium ($x<0$) by \cite{Matloob1996}
\begin{align}
&\hspace{-0.4cm}G_{\{x<0\}}(x,\omega,x')\nonumber\\
  =\;&\frac{i\,\theta(-x')}{2k_1}\Big(e^{ik_1|x-x'|}+\mathcal{R}_1e^{-ik_1(x+x')} \Big)\nonumber\\
  &+\frac{i\,[\theta(x')-\theta(x'-d)]}{2k_2}\mathcal{T}_1'e^{-ik_1x}\Big(e^{ik_2x'}\nonumber\\
  &+\nu\mathcal{R}_2[e^{-ik_2(x'-2d)}+\mathcal{R}_1'e^{ik_2(x'+2d)}]\Big)\nonumber\\
  &+\frac{i\,\theta(x'-d)}{2k_3}\mathcal{T}_2'\mathcal{T}_1'e^{-i[k_1x-k_2d-k_3(x'-d)]}.
 \label{eq:greenslab1}
\end{align}
The first term describes the field generated by the left half-space, the second term
corresponds to the field generated by the medium within the interfaces, and the last
term corresponds to the field generated by the right half-space.
In our notation $\mathcal{R}_1$, $\mathcal{R}_2$,
$\mathcal{T}_1$, and $\mathcal{T}_2$ are the reflection and transmission
coefficients for the first and second interfaces of the two-interface structure.
They are given in terms of the single interface Fresnel coefficients
in Eq.~\eqref{eq:fresnel} as
\begin{align}
 \mathcal{R}_1 & =\frac{r_1+r_2e^{2ik_2d}}{1+r_1r_2e^{2ik_2d}},\hspace{0.5cm}
 \mathcal{R}_2=r_2,\\
 \mathcal{T}_1 & =\frac{t_1}{1+r_1r_2e^{2ik_2d}},\hspace{0.5cm}
 \mathcal{T}_2=t_2.
\end{align}
The parameter $\nu$ in the Green's function is expressed as
\begin{equation}
 \nu=\frac{1}{1+r_1r_2e^{2ik_2d}}.
\end{equation}
The Green's function in the right medium ($x>d$) is, respectively, given by \cite{Matloob1996}
\begin{align}
&\hspace{-0.4cm}G_{\{x>d\}}(x,\omega,x')\nonumber\\
  =\;&\frac{i\,\theta(-x')}{2k_1}\mathcal{T}_1\mathcal{T}_2e^{i[k_3(x-d)+k_2d-k_1x']}\nonumber\\
  &+\frac{i\,[\theta(x')-\theta(x'-d)]}{2k_2}\mathcal{T}_2e^{ik_3(x-d)}\Big(e^{ik_2(d-x')}\nonumber\\
  &+\nu\mathcal{R}_1'[e^{ik_2(x'+d)}+\mathcal{R}_2e^{ik_2(3d-x')}]\Big)\nonumber\\
  &+\frac{i\,\theta(x'-d)}{2k_3}\Big(e^{ik_3|x-x'|}+\mathcal{R}_2'e^{ik_3(x+x'-2d)}\Big).
 \label{eq:greenslab3}
\end{align}
and inside the cavity ($0<x<d$), the Green's function is expressed as
\begin{align}
&\hspace{-0.4cm}G_{\{0<x<d\}}(x,\omega,x')\nonumber\\
  =\;&\frac{i\,\theta(-x')}{2k_1}\mathcal{T}_1\Big(e^{i(k_2x-k_1x')}+\mathcal{R}_2e^{-i[k_2(x-2d)+k_1x']}\Big)\nonumber\\
  &+\frac{i\,[\theta(x')-\theta(x'-d)]}{2k_2}\Big(e^{ik_2|x-x'|}\nonumber\\
  &+\nu\mathcal{R}_2[e^{-ik_2(x+x'-2d)}+\mathcal{R}_1'e^{-ik_2(x-x'-2d)}]\nonumber\\
  &+\nu\mathcal{R}_1'[e^{ik_2(x+x')}+\mathcal{R}_2e^{ik_2(x-x'+2d)}]\Big)\nonumber\\
  &+\frac{i\,\theta(x'-d)}{2k_3}\mathcal{T}_2'\Big(e^{-i[k_2(x-d)-k_3(x'-d)]}\nonumber\\
  &+\mathcal{R}_1'e^{i[k_2(x+d)+k_3(x'-d)]}\Big).
 \label{eq:greenslab2}
\end{align}
Again, the three different terms describe the fields generated by different source domains.

\end{document}